\documentclass[12pt,preprint]{aastex}
\usepackage{color}

\newcommand{\MC}{\multicolumn}
\newcommand{\ESO}{ESO\,474$-$G040}

\DeclareRobustCommand{\ion}[2]{%
\relax\ifmmode
\ifx\testbx\f
{\mathrm{#1\,\textsc{#2}}}\else
{\mathrm{#1\,\mathsc{#2}}}\fi
\else\textup{#1\,{\mdseries\textsc{#2}}}%
\fi}

\begin{document}

\title{The empty ring galaxy  ESO 474 - G040}

\author{Noah Brosch\altaffilmark{}}
\affil{The Wise Observatory and the Raymond and Beverly Sackler School of Physics and
Astronomy, the Faculty of Exact
Sciences, \\ Tel Aviv University, Tel Aviv 69978, Israel}
\email{noah@wise.tau.ac.il}

\author{Petri V\"{a}is\"{a}nen\altaffilmark{}}
\affil{South African Astronomical Observatory, PO Box 9, Observatory, Cape Town, 7935, South Africa, \\ and Southern African Large Telescope Foundation, PO Box 9, 7935 Observatory, Cape Town, South Africa}
\email{petri@saao.ac.za}

\author{Alexei Y. Kniazev\altaffilmark{}}
\affil{South African Astronomical Observatory, PO Box 9, Observatory, Cape Town, 7935, South Africa, \\ and Southern African Large Telescope Foundation, PO Box 9, 7935 Observatory, Cape Town, South Africa, \\ and Sternberg Astronomical Institute of Lomonosov Moscow State University, Moscow 119992, Russia}
\email{akniazev@saao.ac.za}

\author{Alexei Moiseev\altaffilmark{}}
\affil{Special Astrophysical Observatory, Russian Academy of Sciences, Nizhniy Arkhyz, Karachai-Cherkessian Republic 357147, \\ and Sternberg Astronomical Institute of Lomonosov Moscow State University, Moscow 119992, Russia}
\email{moisav@sao.ru}


\begin{abstract}
We describe observations of the apparently empty ring galaxy ESO 474-G040 obtained with the Southern African Large Telescope (SALT). The observations, consisting of imaging, long-slit spectroscopy and Fabry-P\'{e}rot mapping of the H$\alpha$ line, allow determining the ring kinematics as well as estimating the metallicity of the ring and the stellar population composition in its various parts. We propose that the object could best be understood as being the result of a past merger of disk galaxies, which formed a gas ring that subsequently disrupted via the bead instability and is presently forming stars. 
\end{abstract}



\section{Introduction}
\label{sec:intro}

Galaxies with various kinds of rings have been recognized in the history of astronomy, and even our Milky Way and M31 have been claimed at times to have external rings (e.g. Koch \& Grebel 2006a, b). An early effort at classifying ring galaxies (RGs), by Theys \& Spiegel (1976, TS7,) subdivided RGs into RE, RN or RK subclasses: empty rings, rings with nuclei, or showing dominant knots. This was revised again  by Few \& Madore (1986, FM86) based on the appearance of the ring: O with a central nucleus, and P where the nucleus is not necessarily central and the ring is knotted, warped or asymmetric. The latter type were found also to have an excess of companions within two ring diameters, interpreted by Few \& Madore as supporting evidence that the rings were formed by galaxy collisions.

A more recent RG reclassification, by Fa\'{u}ndez-Abans \& de Oliveira-Abans (1998, FAOA98) divided RGs into five types based on the ring and bulge morphologies: P (polar), HL (Hoag like), E (elliptical), I (irregular) and CS (centrally smooth). Among the E types, some were classified as RN by TS76 and as P by FM86. This type was further divided into (a)=knotted, (b)=smooth, and (c)=solitaire. The Ea type show ``knotted rings and may be evidence of ring deformation as a result of tidal interaction; they may be undergoing star formation''. The Eb and Ec types have bulges, according to Fa\'{u}ndez-Abans \& de Oliveira-Abans, in contrast to the CS types that do not have bulges but have knotted rings.

One example of an outstanding ringed galaxy is Hoag's Object (HO: Hoag 1950; Finkelman, Moiseev, Brosch \& Katkov  2011; Brosch et al. 2013) where an almost perfect ring surrounds an elliptical galaxy that is also a fast rotator. Brosch (1985) suggested that HO's central object could be the remains of a bar that drove the formation of the ring, more or less in the way proposed by Buta \& Combes (1996), with the bar subsequently relaxed back into the disk. However, this proposal could not be sustained and later  HO was assumed to be the result of external gas accretion, perhaps directly from the cosmic web or via an interaction with a gas-rich galaxy, possibly in the form of a ``polar ring''. Among rings around early-type galaxies (ETGs), such polar rings are of great interest.

Ringed early-type galaxies (R-ETGs) are a rare and important class of objects through which one can investigate both the dark matter contents of galaxies and their haloes (e.g., Khoperskov, Moiseev, Khoperskov, Saburova 2014), and the mechanism by which ETGs are rejuvenated. In the cases studied so far the ETGs show a definite equatorial plane along the major axis, but the ring is generally inclined to this plane implying that it is a different dynamical entity. Rings around ETGs are generally bluer than the cores and often show star formation (SF) exhibited as line emission.

A trending scenario for the formation of ringed ETGs involves accretion of fresh gas and dust (ISM), which is subsequently used for SF. 
 This process could have been relatively mild, with the ISM being accreted via a tidal interaction, or quite violent through a direct collision of the two galaxies. In some cases it is possible to identify a candidate projectile galaxy that is quite close in projected distance and redshift to the R-ETG and which may even be connected with it via stellar and/or gaseous bridges. Sometimes these rings are connected with internal disk structures, such as Lindblad resonances.

Arp (1966) recognized that some peculiar types of ring galaxies did not fit the accepted morphological classifications. Among the objects depicted in his atlas, numbers 146 and 147 stand out; these are luminous rings whose centres are empty, although there are other galaxies very nearby. The empty ring galaxies were interpreted as collisional remains from an encounter between a disk galaxy and another galaxy (Lynds \& Toomre 1976; Appleton \& Struck-Marcel 1996). Collisional RGs may form via a $\sim$head-on passage of a projectile galaxy through a target disk galaxy that is gas-rich. In the Appleton \& Struck-Marcel scenario the ``bull's eye'' collision drives symmetrical density waves into the disk, causes the formation of young stars in a circular pattern, thus visible rings are produced.

Most young stars in collisional RGs are in the rings, which show a moderately high star formation rates (e.g. 18 M$_{\odot}$ yr$^{−1}$ in the Cartwheel galaxy, a prototype galaxy for this class of objects; Mayya et al. 2005).  Numerical simulations showed that the collisional rings are expanding. This was also found observationally {\bf (e.g. Fosbury \& Hawarden 1977 for the Cartwheel, Few, Madore \& Arp 1982 for AM 064-741)}. This was confirmed by HI and/or H$\alpha$ observations of the neutral and ionized gas kinematics, as indeed found from the H$\alpha$ mapping of the Cartwheel galaxy by Amram, Mendes de Oliveira, Boulesteix \& Balkowski (1998) where the ring was found to expand with (13--30)$\pm$10 km s$^{-1}$ or from the study of the ionized gas kinematics in Arp 10, which hosts a ring expanding at 30--110 km s$^{-1}$ (Biziaev, Moiseev \& Vorobyov 2007).

The collisional formation scenario predicts that a ``projectile'' would always be found relatively nearby, since a newly formed ring galaxy is unlikely to retain its shape for significantly longer than about one dynamical time (a few 10$^8$ years) unless it is stabilized by a massive central object or by  a dark mater halo. 
 Even in this case, as shown by Theys \& Spiegel (1976), the ring might fragment into a number of clumps on a similar time scale. Also, N-body simulations by Wu \& Jiang (2012, 2015) argued that, for specific parameters of the collision, an ``intruder'' galaxy might merge with the nucleus of a target galaxy.

Tutukov \& Fedorova (2006) investigated changes in the structure of disk galaxies following a close encounter with another galaxy. They showed that such encounters can excite the formation of inner structures, including spiral arms and even ring structures. However, in this case the central regions of the target galaxy should not be empty. Moreover, collisions and mergers of disk galaxies can give rise, in certain situations, to ring-like structures formed by tidal interactions and compression-driven star formation, such as witnessed in parts of the Antennae (Arp 244) system.

Very recently, {\bf Moreno (2015) and Moreno et al. (2015)} described results of smoothed-particle hydrodynamic simulations of disk galaxy mergers. His results show that mergers between disks with either aligned or perpendicular spins result in all cases in the formation of an off-nuclear gas ring produced by the warping of the tidal tails. This structure manifests itself as a stable star-forming ring that appears 0.2--1 Gyr following the first pericentre passage and ``survives for a prolonged period''.

Finally, RGs and similar objects are not relegated to the nearby Universe. RGs at high redshifts have been studied by Elmegreen \& Elmegreen (2006), based on HST imaging of the GEMS and GOODS fields out to z$\simeq$1.4. They found objects that resemble arcs in partial RGs but lack evident disks, and are clumpy. The clumps in these objects are bluer at all redshifts than the clumps in  RGs and their ages are 10$^7$-10$^8$ yr.

To address the issue of origin via kinematics and  stellar populations in both rings and cores (or projectiles) we selected a small number of R-ETGs from the recent catalog of Moiseev et al. (2011) from which we chose objects most similar in appearance to Hoag's Object, and a number of core-less ring galaxies from the Arp et al. (1987) ``Catalogue of southern peculiar galaxies and associations'' that have a similar appearance (Category 6a: ``Empty rings and displaced nuclei'').

\begin{figure}[th]
\centering{
  \includegraphics[width=16cm]{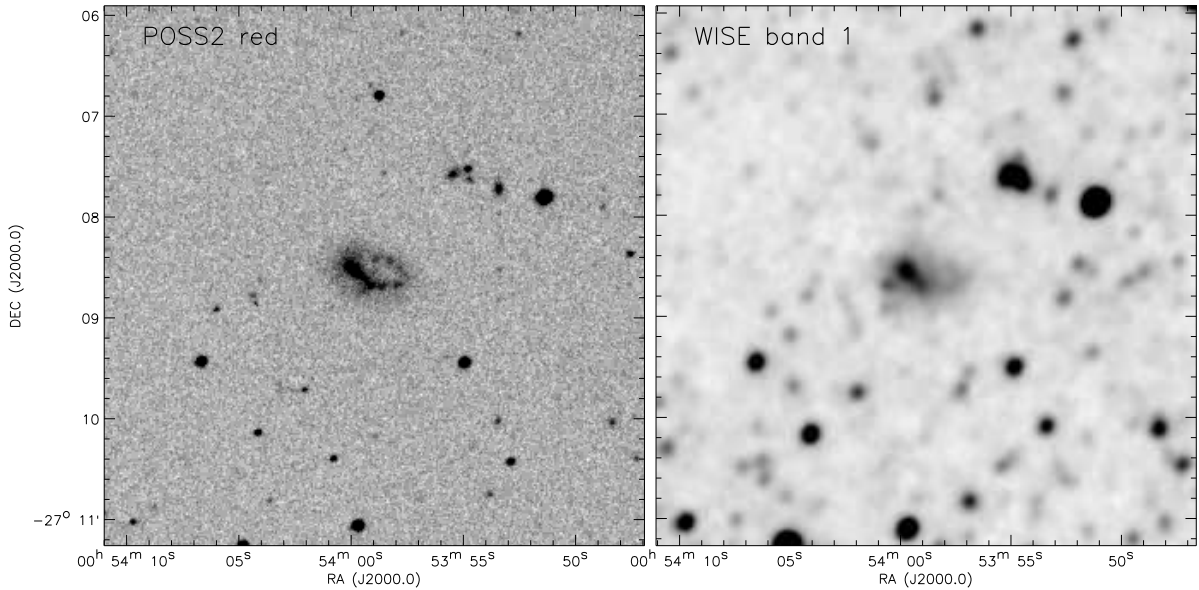}
  \caption{Images of ESO 474-G040 (north is up and East is to the left). {\it Left panel:} Red image from the CADC archives shows a beaded ring appearance with an apparently empty interior. {\it Right panel:} Image in the WISE W1 band (3.4 $\mu$m) shows a similar structure.}
  \label{fig:FC}}
\end{figure}

The galaxy studied here is one of the objects in this latter list: ESO 474-G040 (hereinafter ESO474), an empty ring galaxy corresponding to the class RE (elliptical rings with empty interiors) in the TS76 classification, or possibly Ea in the FAOA98 scheme. This galaxy is listed as an Irr in the UGC (Lauberts 1982) and as a spiral in the APM catalog (Loveday 1996). It has two discordant redshifts in NED\footnote{NASA/IPAC Extragalactic Database at http://ned.ipac.caltech.edu/}; 29200$\pm$89 km s$^{-1}$ from the 2dF final data release and 6101 km s$^{-1}$ 
 or 5996$\pm$300  km s$^{-1}$ (in ADS, listed as private communication from M. Colless and originating from the AUTOFIB redshift survey). The object was listed as AM 0051-272 among the peculiar galaxies of Arp \& Madore (1987), and was classified there as a pair, with one of the galaxies being a ring. The foreground extinction, from NED, is A$_R$=0.033 mag. A sky survey R-band image of the object is shown in the left panel of Figure~\ref{fig:FC}.

The structure of this paper is as follows: in Section~\ref{sec.obs} we describe our observations and their reductions. Section~\ref{sec.proc} details the analyses of the reduced data and the results of these analyses are described. The results are discussed in Section~\ref{sec.discuss} and are put in the context of other ringed galaxies. Section~\ref{sec.summary} summarizes this paper.

\section {Observations and data reduction}
\label{sec.obs}
We observed ESO474 using the Southern African Large Telescope (SALT; O'Donoghue et al. 2006, Buckley et al. 2006) using the Robert Stobie Spectrograph (RSS; Burgh et al. 2003) in two modes: long-slit classical spectroscopy (LS) and imaging Fabry-P\'{e}rot interferometer (FP) mode.  The goal of the first type of observation was to characterize the light emitted from the ring through the continuum and the emission lines, and to derive the proper radial velocity as well as the physical conditions in various regions of the ring. The FP data were used to derive the two-dimensional kinematics of the ionized gas in the ring via the H$\alpha$ emission.

\begin{figure}
\centering{
  \includegraphics[width=16cm]{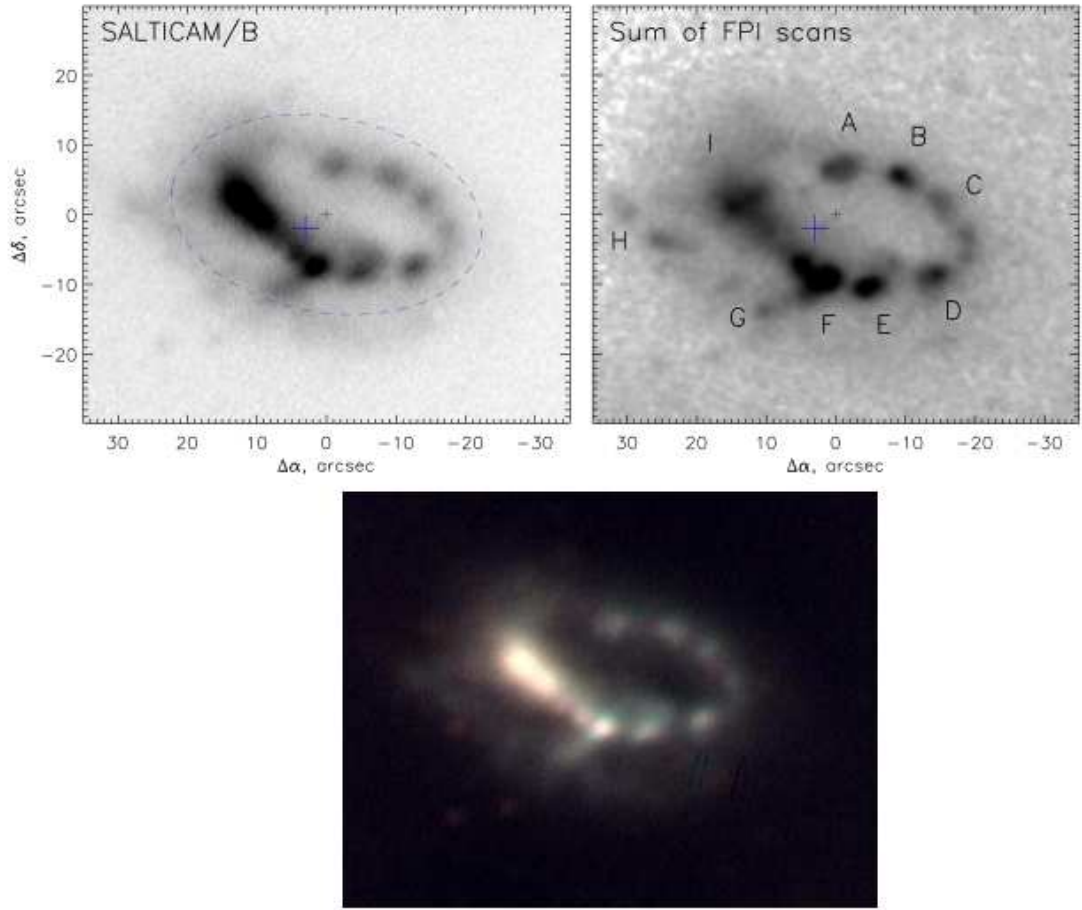}
  \caption{Deep images of ESO474-G040. {\it Upper left:} A 4 min SALTICAM B-band exposure of ESO474. The dashed ellipse shows an approximation to the ring's external isophotes. {\it Upper right:} A deep H$\alpha$+continuum image of the galaxy obtained as the sum of 56 FP scan steps of approximately 3 min each, without flux or seeing corrections.  Knots referred to in text are labelled. {\it Lower panel:} An RGB image using the SALTICAM r', g' and B-band images.} }
  \label{fig:474b}
\end{figure}

We also collected fairly deep images of the object in the Johnson B-band and the SDSS $g'$ and $r'$ bands
using the SALTICAM imager on SALT. These combine two frames per filter each exposed for 120 s, and were used to determine the morphology of the object at fainter brightness levels than allowed by images in public data bases.  The B-band image is shown in the left panel of  Figure~\ref{fig:474b} and a 3-color version at bottom.  We do not attempt a photometric calibration of these images.

A single preliminary LS spectrum was taken on 17 Nov 2012.
This spectrum immediately indicated the necessity of a downward revision of the NED redshift from z=0.098 to 0.019. Based on this new determination, confirming the lower redshift entry in NED, four further LS spectra were taken at different Position Angles (PA) 
during the next observing semesters between 09 Sep 2013 and 21 Jun 2014.  All the LS spectra had a total exposure time of 40 min broken up to 3 x 800s individual frames.  The slit width used was 1.25 arcsec and, with the grating used, the spectra cover the range 3750\AA\, to 6900\AA\, with a resolution of $\sim$4.5 \AA\,. The preliminary spectrum taken in 2012 covers the range 4600-7600\AA.

We obtained five FP data sets for the galaxy between 30 Jun 2014 and 01 Sep 2014 using the RSS in a single
etalon medium resolution (MR) mode, stepped by 5\AA\, between each velocity slice of the data cube.  The SALT RSS/FP system has been described by Rangwala, Williams, Pietraszewski \& Joseph  (2008), and see also Mitchell et al. (2015) for recent status.
 At $\sim$6700\AA\,, corresponding to the galaxy's H$\alpha$ line, the FP offers a resolution of R=1250 and we chose 4x4 spatial binning where each pixel corresponds to $\approx$0.5 arcsec. Most of the FP sets consisted of thirteen scans 
 each exposed for 183 s. Not all data sets could be used for the analysis presented here; two were rejected from the kinematic analysis described below because the FP plates showed bad parallelism that increased significantly the line width. Nevertheless,  all these frames (56 in total) were combined to construct a stacked deep H$\alpha$+continuum image of the galaxy which has a final seeing of about 2.2 arcsec. This image is shown in the right panel of Figure~\ref{fig:474b}.

\begin{figure}
\centerline{
\includegraphics[width=8 cm]{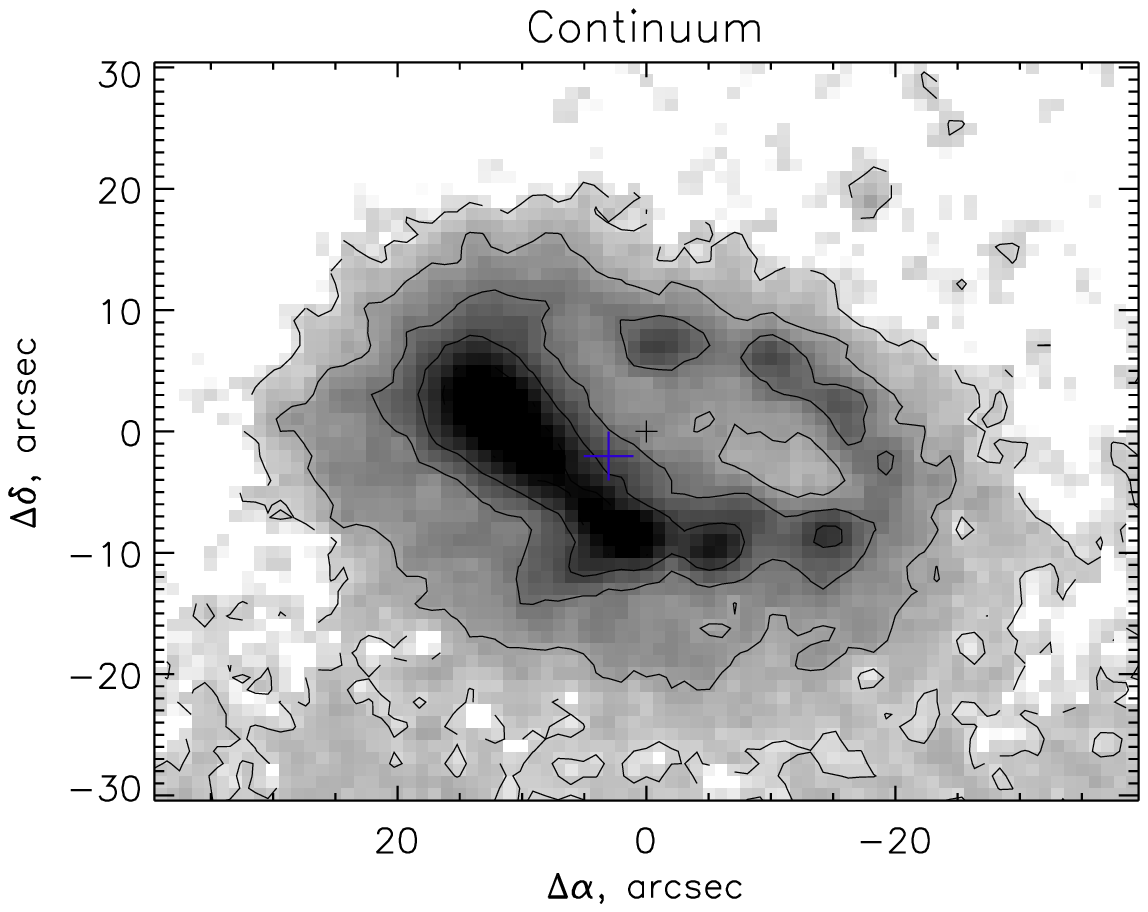}
\includegraphics[width=8 cm]{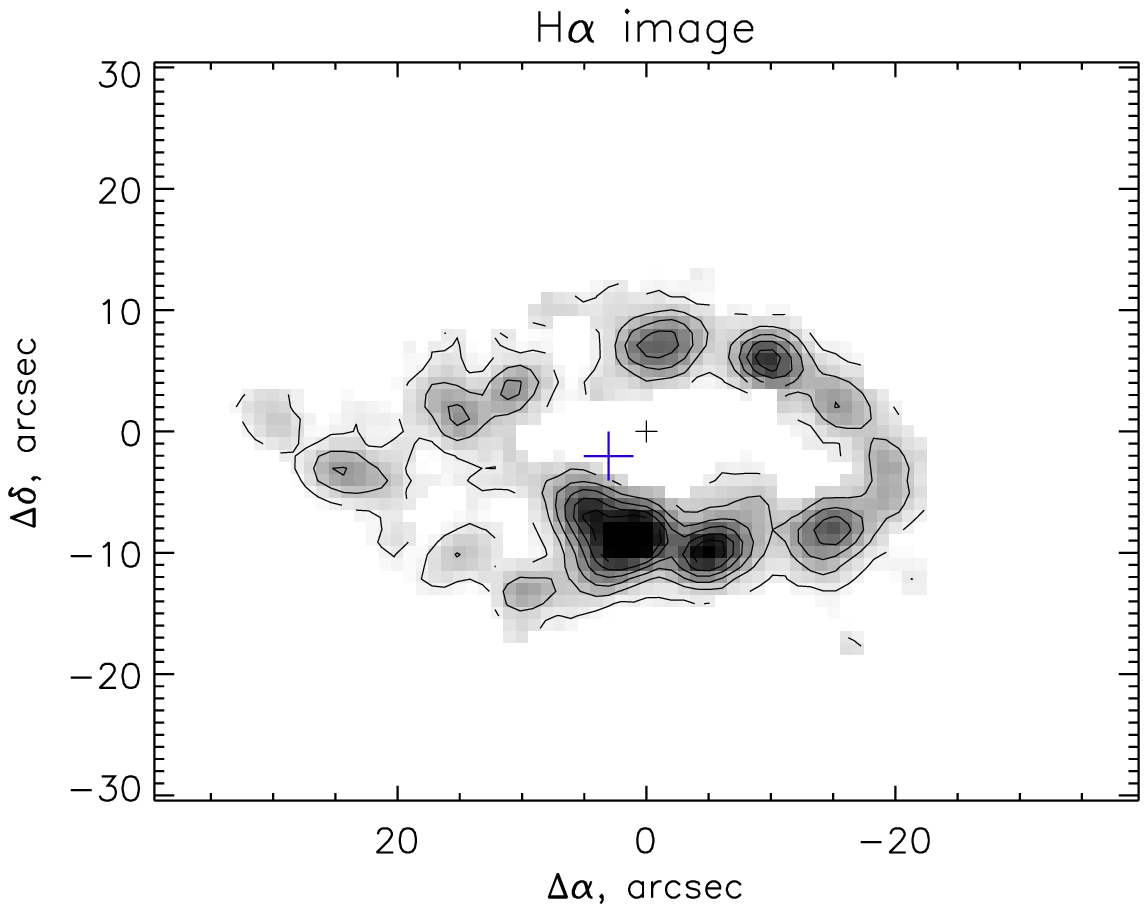}}
\centerline{\includegraphics[width=8 cm]{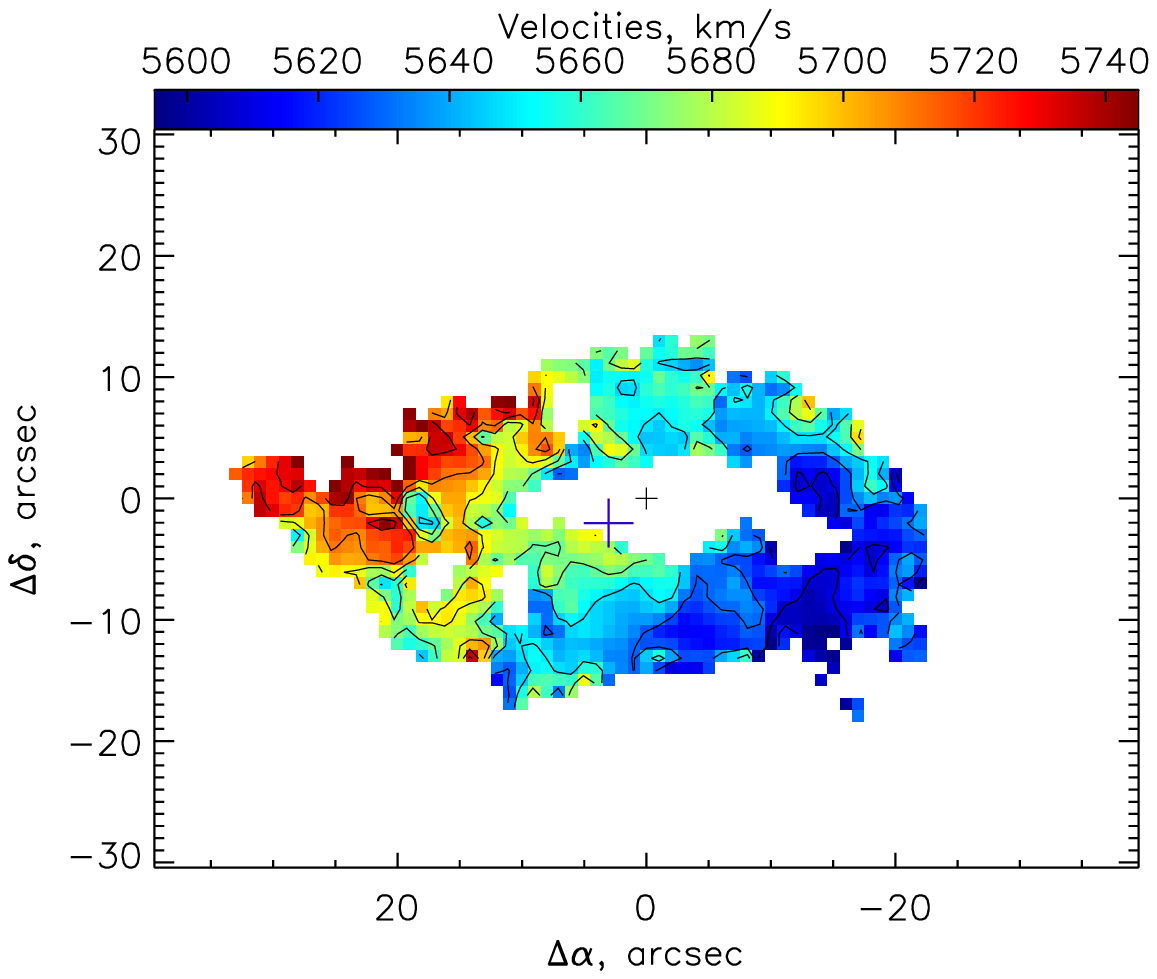}
\includegraphics[width=8 cm]{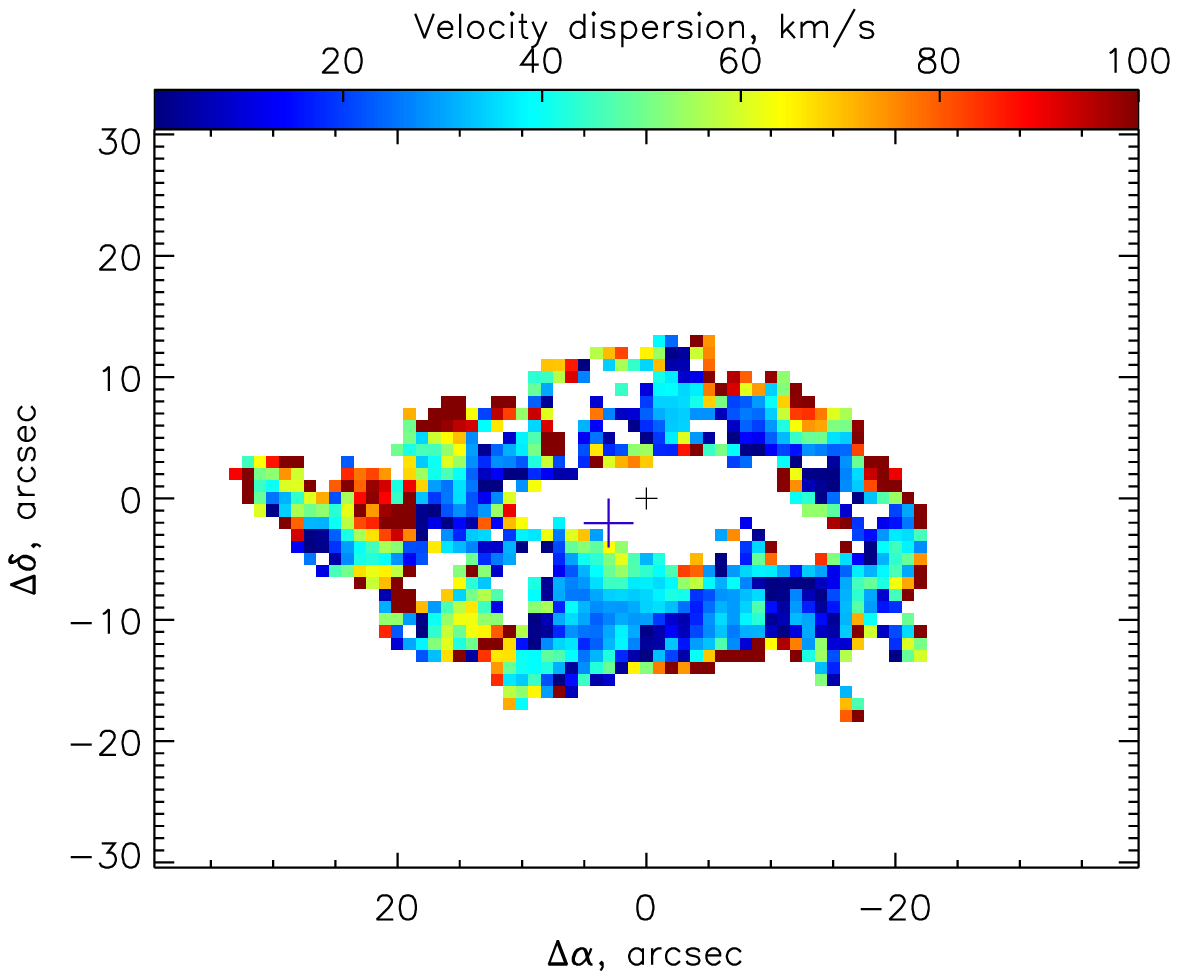}}
\caption{ESO 474-G040 maps derived from fitting Voigt-profiles to the FP data cube, combining three data sets. Top row: image in the continuum (left) and in the H$\alpha$ emission line (right). Bottom row: the line-of sight  velocity field (left) and  velocity dispersion  (right). {\bf On these and forthcoming figures the smaller cross at (0,0) marks the photometric centre of the ring, while the larger blue cross corresponds to the kinematic centre. The $\alpha$ and $\delta$ offsets are given relative the photometric centre whose coordinates are given in  section 3.4.}}
\label{fig:Moiseev_FPI}
\end{figure}

Since the FP provides essentially a velocity for each location in the image, a velocity map shown in the bottom-left panel of Figure~\ref{fig:Moiseev_FPI} was produced along with a velocity dispersion map shown in the bottom-right panel of the same figure. For each location, a Voigt profile was fitted to the line intensity vs. the scan velocity and the peak was selected as representing the ionized gas velocity at that location.
We identified nine bright knots in the stacked FP image; these are labeled A through I in  Figure~\ref{fig:474b}. 

\begin{figure}[th]
\centering{
  \includegraphics[width=15cm]{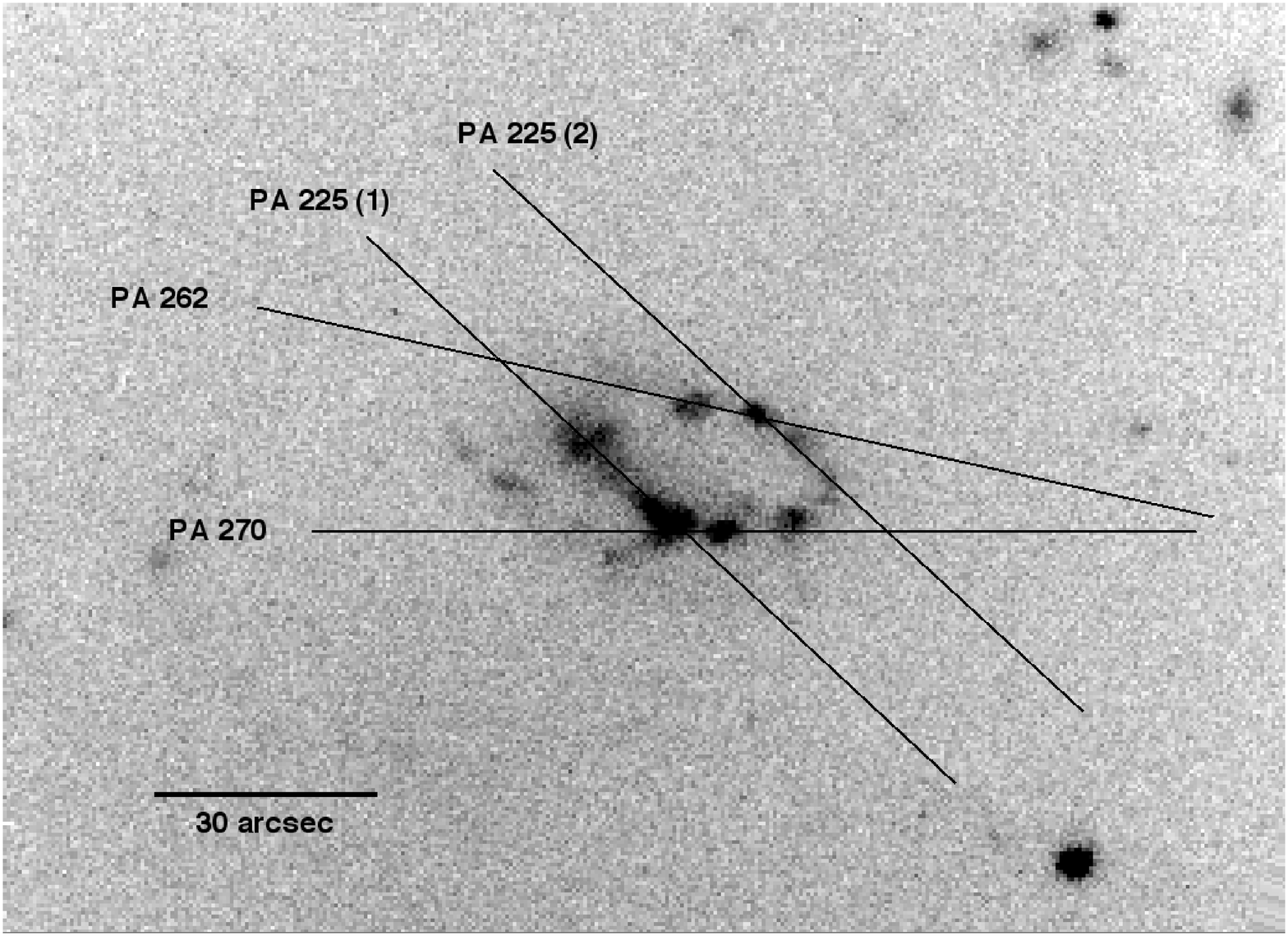}
  \caption{ESO 474-G040 sum of FP scans with the representation of the four slit positions for LS spectroscopy.}
  \label{fig:fc_slits}}
\end{figure}



\begin{figure}
\centering{
  \includegraphics[width=16cm]{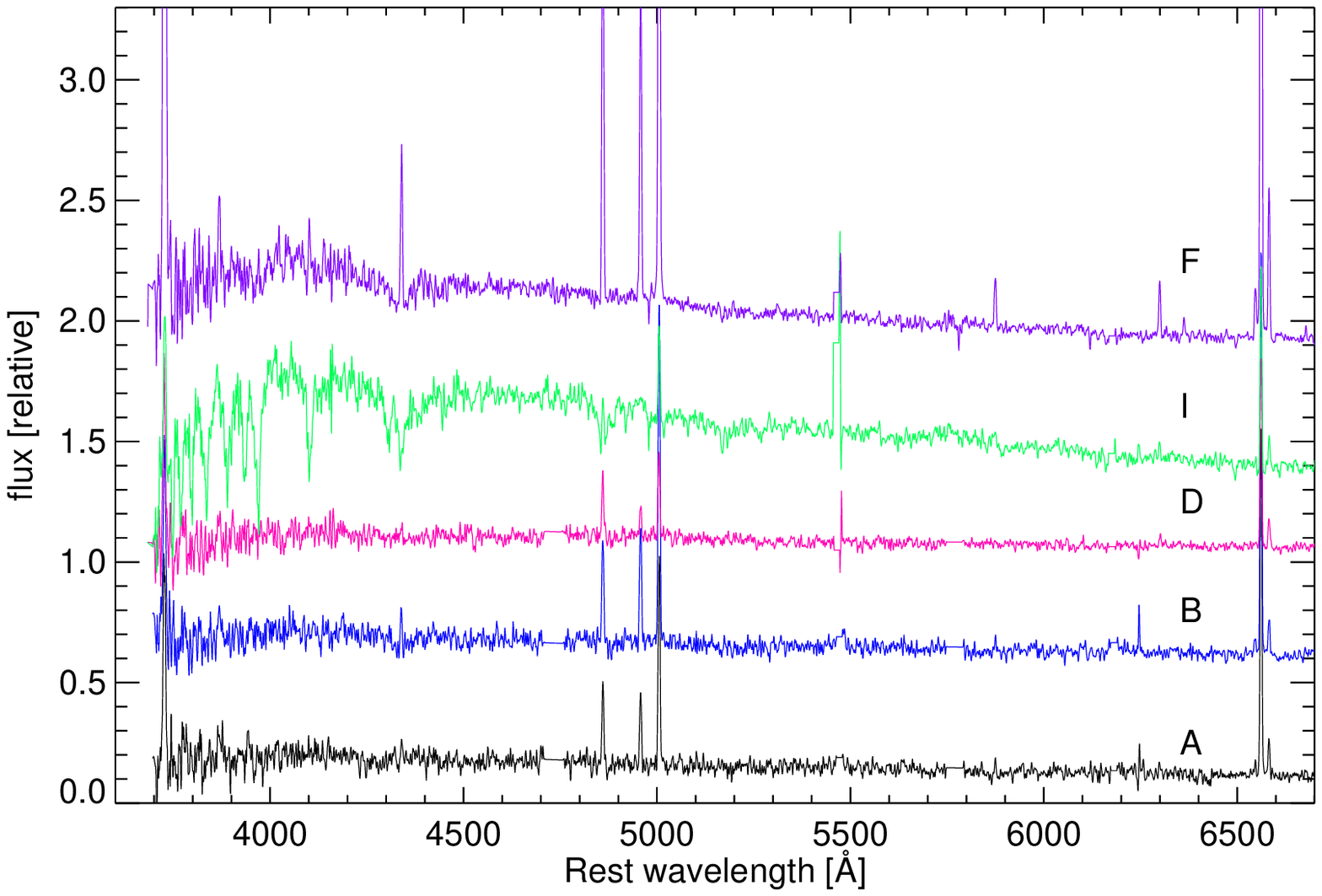}
  \caption{Spectra of the knots A, B, D, F, and I starting from the bottom. Note that while knot I is the brightest in continuum, knot F shows much stronger emission lines. The plot is in relative energy units and has been shifted to  rest wavelength.}
  \label{fig:knots_FI}}
\end{figure}

The slit positions for the LS observations, shown in Figure~\ref{fig:fc_slits}, were selected to sample as many of these knots as possible.
The long-slit spectra were taken at three position angles: 225$^{\circ}$ (two parallel spectra sampling different locations of the galaxy), 262$^{\circ}$, and 270$^{\circ}$.  The first spectrum at PA=225$^{\circ}$ passes through knots F and I.  The second spectrum at the same PA samples knots B and C. The one at PA=270$^{\circ}$ examines knots D, E and F; the fourth spectrum, in PA=262$^{\circ}$, checks knots A and B. Therefore, two knots, B and F, have two spectra each, which allows for cross-comparison and yields a slightly better S/N by combining the two spectra. The analysis below considers the combined spectra for these two knots.  Each long-slit spectrum was wavelength-calibrated and spatially reduced.  Those spectra which resulted in satisfactory 1D extractions are shown in Figure~\ref{fig:knots_FI}.

Note that the brightest knot (I) in broad band is {\em not} the brightest in line emission; the strongest H$\alpha$ knot is F as seen in the line-only image (compare top panels of Fig.~\ref{fig:Moiseev_FPI}).
Consistently, the more intense line emission comes from the F knot, whereas the brighter I knot contributes mostly continuum and shows only a faint emission line component. 

\section {Analysis and results}
\label{sec.proc}

\subsection{General information}

Table 1 collects general data about ESO474 gathered from public data sets. The WISE photometry is explained in the on-line mission document WISE/IPAC.
 The magnitudes in the different bands are derived from the background-subtracted counts in digital units as:
\begin{equation}
M_{cal}=M_{0, inst}-2.5 log_{10}(counts)
\end{equation}
with M$_{0, inst}$ being 20.730 for W1 and 19.567 for W2. The AllWISE Source Catalog at IPAC\footnote{irsa.ipac.caltech.edu} breaks ESO474 into four sources separated by a few arcsec. We performed aperture photometry on the calibrated WISE images and found, for the entire galaxy, the magnitudes listed in the table. We are listing there the magnitudes for only the first two bands, W1 (3.4 $\mu$m) and W2 (4.6 $\mu$m), since the images of the two longer-wavelength bands are weak and are noise-dominated.

\begin{table}
\label{t.obsdata}
\begin{footnotesize}
\begin{center}
\vspace{0.5cm}
\begin{tabular}{|c|c|c|}
\hline
 Parameter & Value & Source  \\
\hline
$\alpha$ (J2000)  & 00$^h53^m59^s.2$ & NED  \\
$\delta$ (J2000)  & {\bf --27$^{\circ}$08'36".8s} & NED  \\
FUV {\bf (mag.)} & 17.40(0.03) & GALEX/NED  \\
NUV  {\bf (mag.)} & 17.19(0.02) & GALEX/NED  \\
U$_T$ {\bf (mag.)} & 15.73(0.10) & M98   \\
B$_T$ {\bf (mag.)} & 15.93(0.10) & M98  \\
V$_T$ {\bf (mag.)} & 15.63(0.10) & M98  \\
R$_T$ {\bf (mag.)} & 15.16(0.06) & M98  \\
I$_T$ {\bf (mag.)} & 14.84(0.10) & M98  \\
3.4 $\mu$m {\bf (mag.)} & 13.58(0.02) & WISE-1 GATOR   \\
4.6 $\mu$m {\bf (mag.)} & 13.54(0.02) & WISE-2 GATOR   \\
\hline
\end{tabular}
\caption{Observational data for ESO 474 - G040}
\end{center}
Notes to Table 1: The numbers in parentheses following the value are the uncertainty in the value. The reference for the various parameters is M98=Metcalfe et al. (1998). The GALEX magnitudes are total, in the AB system, and integrated from map as listed in NED.
\end{footnotesize}
\end{table}

With the revised redshift for ESO474, at a distance of $\sim$78 Mpc, the size of the object ($\sim$56 arcsec, from our deep image shown in the top-left panel of Fig.~\ref{fig:474b}) translates into a major diameter of $\sim$21 kpc and the absolute magnitude of the entire galaxy is about --18.5 mag, fainter than an L$_*$ galaxy but slightly brighter than the absolute magnitude threshold to be considered a dwarf galaxy (about --18 mag).


\subsection{Neighbourhood}

We searched NED for objects within 60 arcmin and 500 km s$^{-1}$ from ESO474. We found the five objects listed in Table 2; these are closer than 1.3 Mpc with the nearest galaxy being slightly closer than 150 kpc in projected distance. All five galaxies are late types and some or all show star formation. The morphological classification given in the last column is from NED, with the exception of one object where this was derived from a visual inspection of the image.
 {\bf Using the search tool provided in NED, we find two galaxies within 500 km s$^{-1}$ and 1 Mpc projected distance; these are the top two in Table 2. Within 2 Mpc and 1000 km s$^{-1}$ NED lists 15 galaxies.}

\begin{table}
\label{t.objects}
\begin{footnotesize}
\begin{center}
\vspace{0.5cm}
\begin{tabular}{|c|c|c|c|c|c|}
\hline
Object Name & $\alpha$ J2000 & $\delta$ J2000 & v$_{\odot}$ (km s$^{-1}$) & m$_V$ {\bf (mag.)}; Morphology & M$_V$ \\
\hline
ESO 474 - G 039               & 00h53m43.4s & --27d02m59s & 5552(10) & 14.9; SA(r)bc HII & --19.43 \\
GALEXASC J005241.40-270541.9  & 00h52m41.4s & --27d05m42s & 5670(64) & 18.5; disk & --16.30 \\
ESO 411 - G 034               & 00h57m46.9s & --27d30m06s & 5588(06) & 13.6; SAB(r)c: & --20.84 \\
6dF J0057469-273006           & 00h57m46.9s & --27d30m06s & 5541(45) & 12.3; HII & --22.10\\
ESO 474 - G 032               & 00h50m50.4s & --26d29m30s & 5514(10) & 16.57; Sc & --17.82 \\
\hline
\end{tabular}
\caption{Objects in the neighbourhood of ESO 474 - G040}
\end{center}
Note to Table 2: The value in parentheses following the velocity is its uncertainty in km s$^{-1}$. Column 5 gives the apparent magnitude of the object and its morpological classification. Column 6 shows the absolute magnitude from NED, after correcting for the revised redshift of ESO474.
\end{footnotesize}
\end{table}

 The nearest galaxy, ESO 474 G039, is an $\sim$edge-on disk that does not show morphological disturbances or signs of interaction; these could have been expected if this object would have been the projectile. The photometric indices from the GALEX FUV to the infrared ones from IRAS are consistent with a star-forming disk. The other galaxies are more than twice as distant from ESO474 and all are consistent with star-forming disks. We conclude that this part of the sky seems to harbor mostly normal star-forming galaxies, but that these do not show obvious signs of interaction nor does any look like being the possible projectile.

\subsection{Morphology}

The morphology of ESO 474 G040 can be better understood from the SALTICAM images and from our  stacked FP data than from publicly-available sky survey images. The images shown in Figs.~\ref{fig:474b} and~\ref{fig:Moiseev_FPI} demonstrate clearly that the object exhibits an apparent ring, with various knots.
The ring is wider and fuzzier in continuum emission than in the H$\alpha$ line emission originating from the knots; the latter form a ring with a smaller diameter, some 30 arcsec=11 kpc with a possible extension to the East. The line emission is concentrated in the central regions of the continuum distribution. Knot I is faint in H$\alpha$ but is bright in the optical continuum. Note, in particular, that even using the deep image data from SALT the centre of the ring remains empty.



The deeper SALTICAM images in Figure~\ref{fig:474b} show that there is extended ``fuzz'' to the south of the ring feature, perhaps best seen in the 3-color version.  
The figure also shows that the fuzz seems to be connected to knot F by a spur to the SE. With some imagination, one could believe that the fuzz forms a second, larger ring, extending perhaps beyond the more luminous main ring, similar to what
one sees in an extended low-surface-brightness (LSB) disk.



Additional information about the object morphology can be derived from the WISE W1 (right panel of Figure~\ref{fig:FC}) and W2 images. Despite the low S/N of the outer parts and the lower angular resolution, with the WISE PSF in W1 being 8".3, it is clear that knot I is the brightest feature and that the ring, which appeared knotted in the bluer spectral bands with the knots A, B, and C, seems here to be $\sim$ uniform. The ring appears ``wider'' to the south, just as the continuum image in the top-left panel of Figure~\ref{fig:Moiseev_FPI} indicates. 

\subsection{Kinematic properties}

The kinematic centre, determined as the centre of symmetry in the velocity distribution, is located at [$\alpha$(2000.0)=00:53:59.1:  $\delta$(2000.0)=-27:08:36.5] with our astrometry solution. However, a formal photometric centre of the elliptical ring lies some 3.5 arcsec NW of the point at [$\alpha$(2000.0)= 00:53:58.8; $\delta$(2000.0)=-27:08:34.5]. If the ring is flat and circular then its line of nodes is oriented to PA=72$^{\circ}\pm3^{\circ}$ with an inclination of  i=52$^{\circ}\pm3^{\circ}$.

We fitted the observed velocity field using tilted-ring models 
 using both positions of the centre of rotation, the kinematic and the photometric. The orientation parameters  derived from  FP velocity field in the model centred to the kinematic centre were
$PA=63^{\circ}\pm4^{\circ}$; $i=48^{\circ}\pm5^{\circ}$; $V_{sys}=5664\pm4$ km s$^{-1}$; for the alternative centre {\bf $PA$ and $i$ were fixed to the above values ($PA=72^{\circ}$, $i=52^{\circ}$), and $V_{sys}$ was found to be 5664$\pm4$ km s$^{-1}$}.

\begin{figure}
\parbox[b]{7cm}{
\includegraphics[width=7 cm]{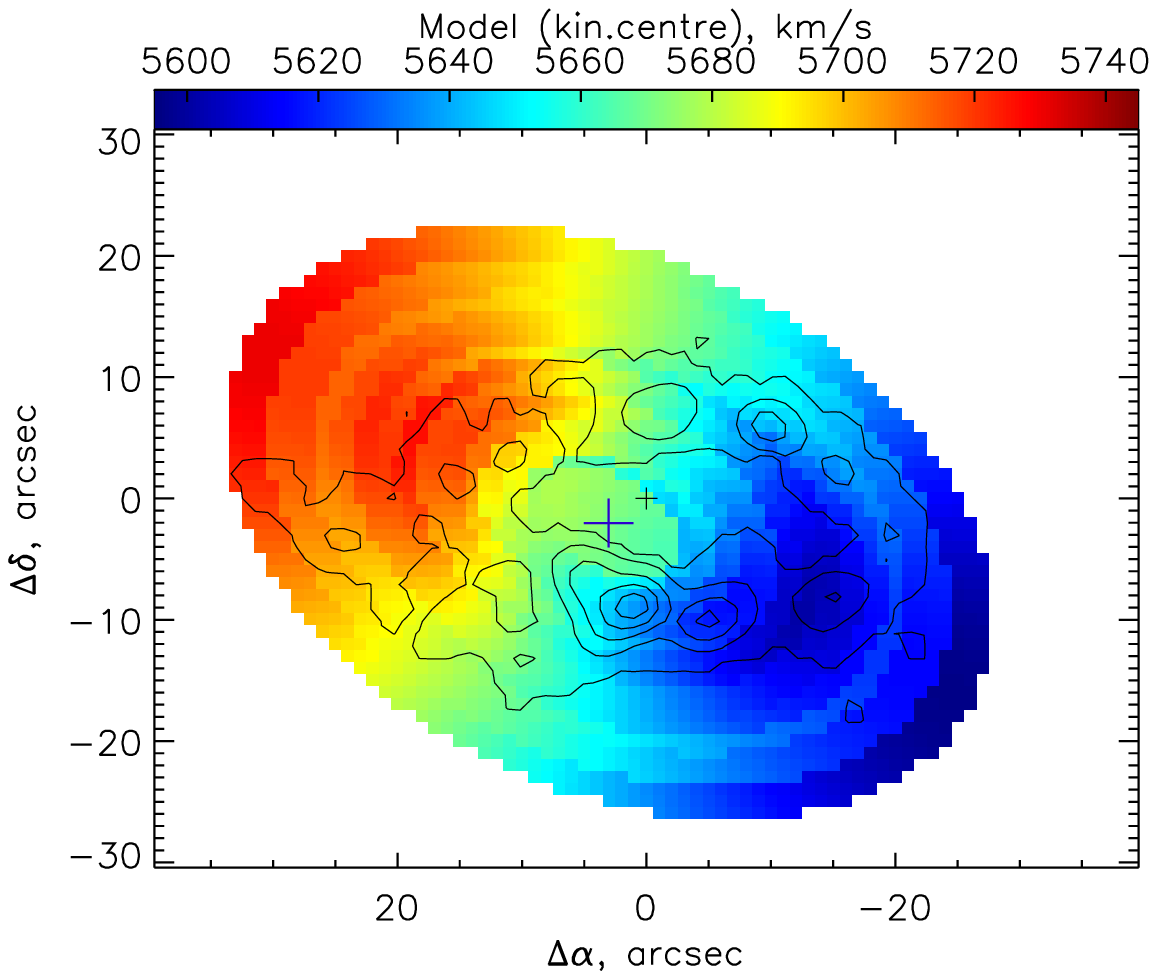}
\includegraphics[width=7 cm]{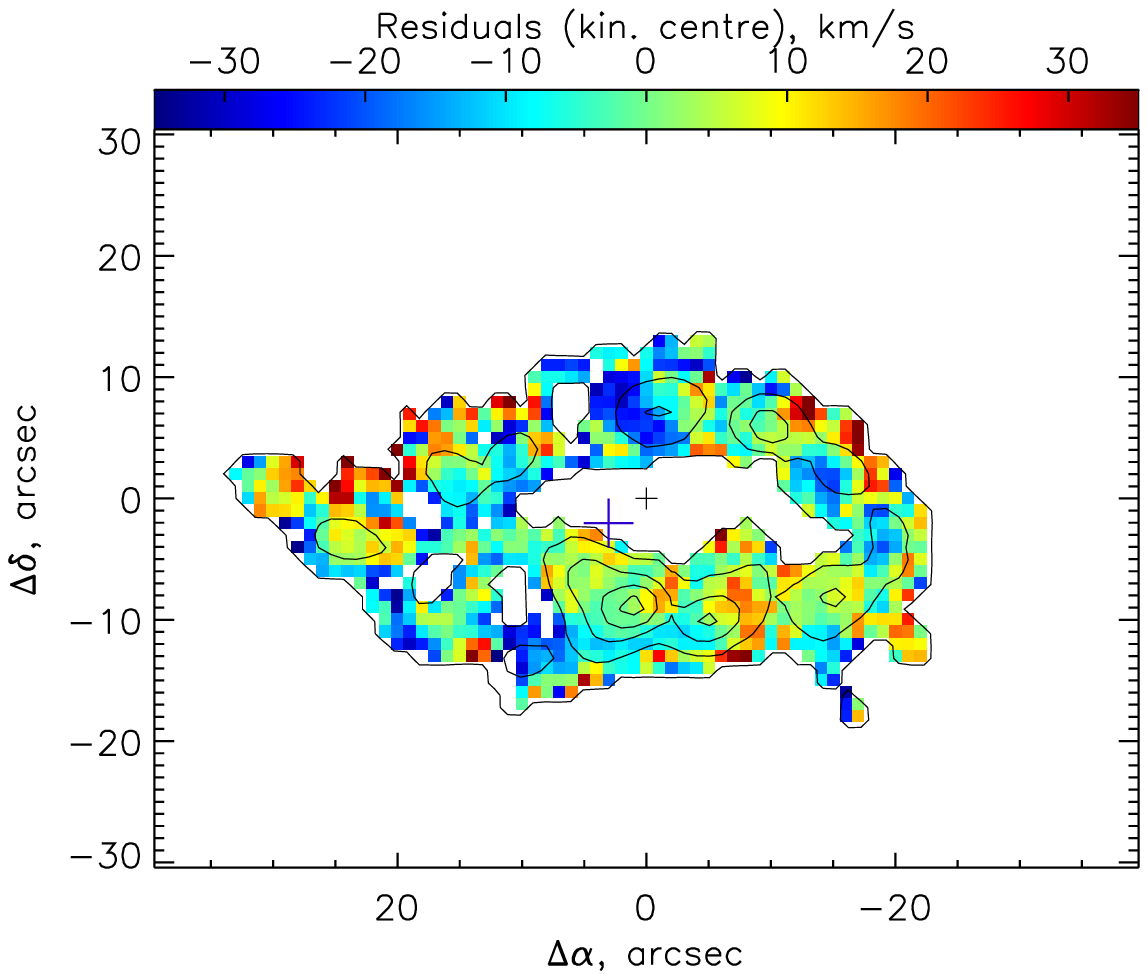}
}
 \includegraphics[width=8 cm]{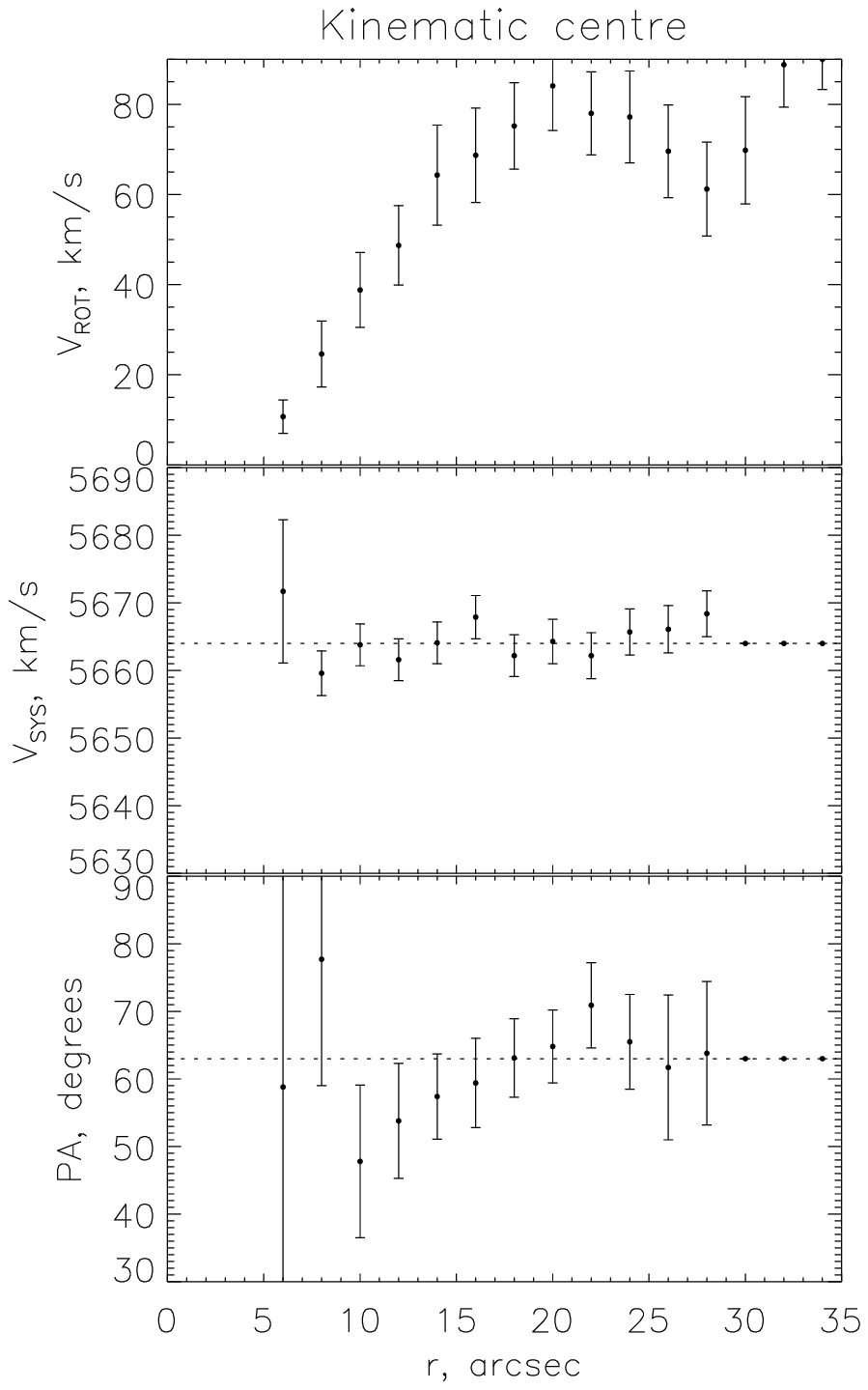}
\caption{Tilted-rings analysis of the velocity field (the accepted rotation centre position coincides with the kinematic centre). Left maps: the model velocity field (top) and residual velocities map. The right panel shows the radial variations of the model parameters. From top to bottom: the circular rotation velocity, the systemic velocity, the line-of-nodes position angle. The dashed line marks the mean values for the galaxy disk.}
\label{fig_mod1}
\end{figure}

\begin{figure}
\parbox[b]{7cm}{
\includegraphics[width=7 cm]{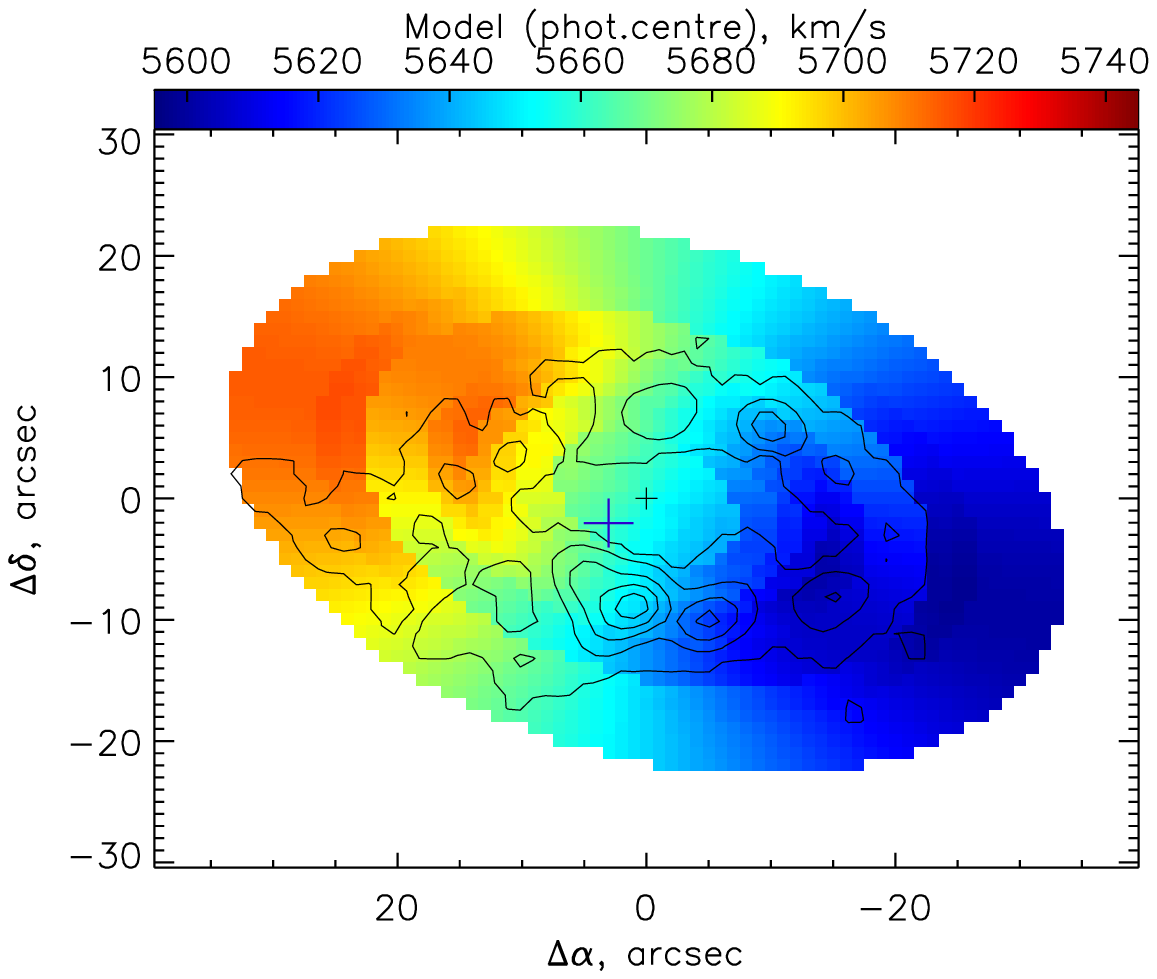}
\includegraphics[width=7 cm]{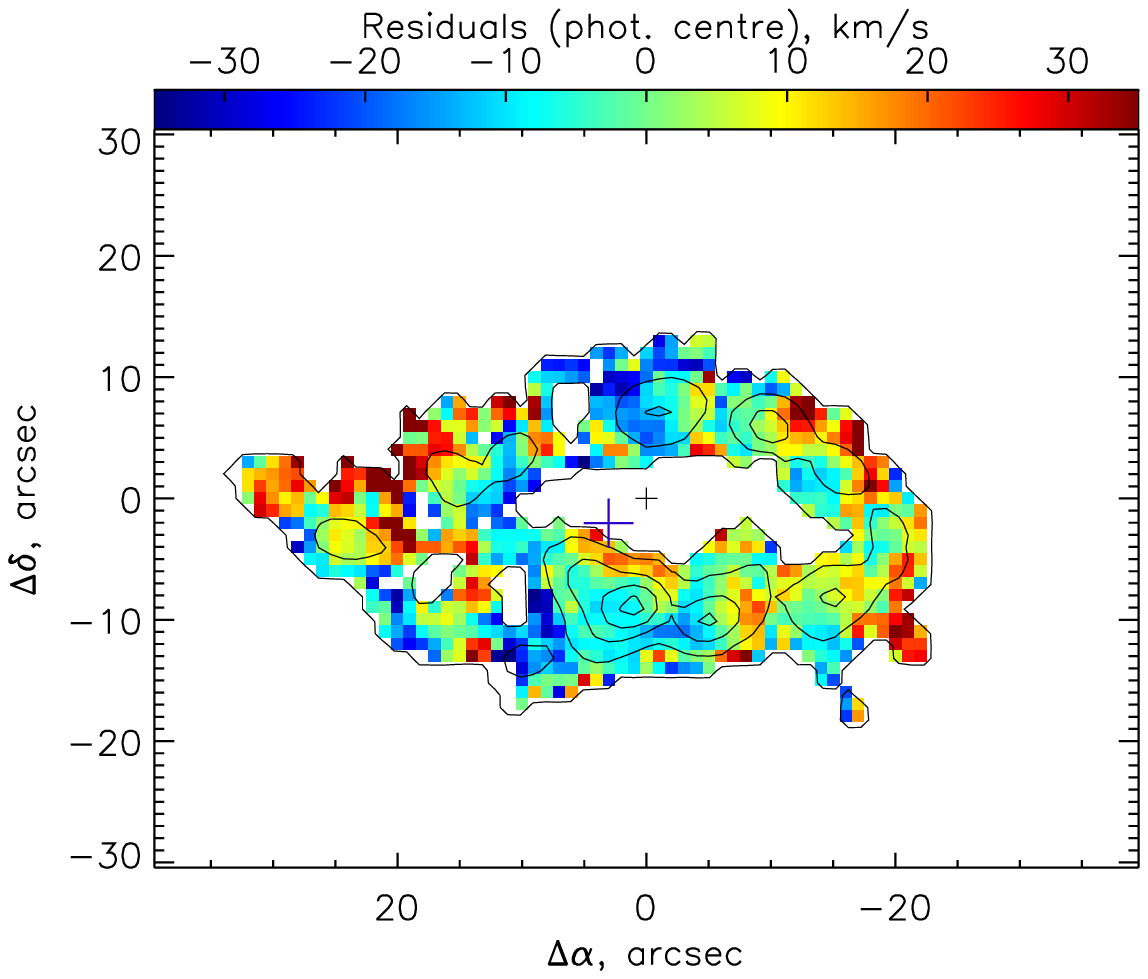}
}
 \includegraphics[width=8 cm]{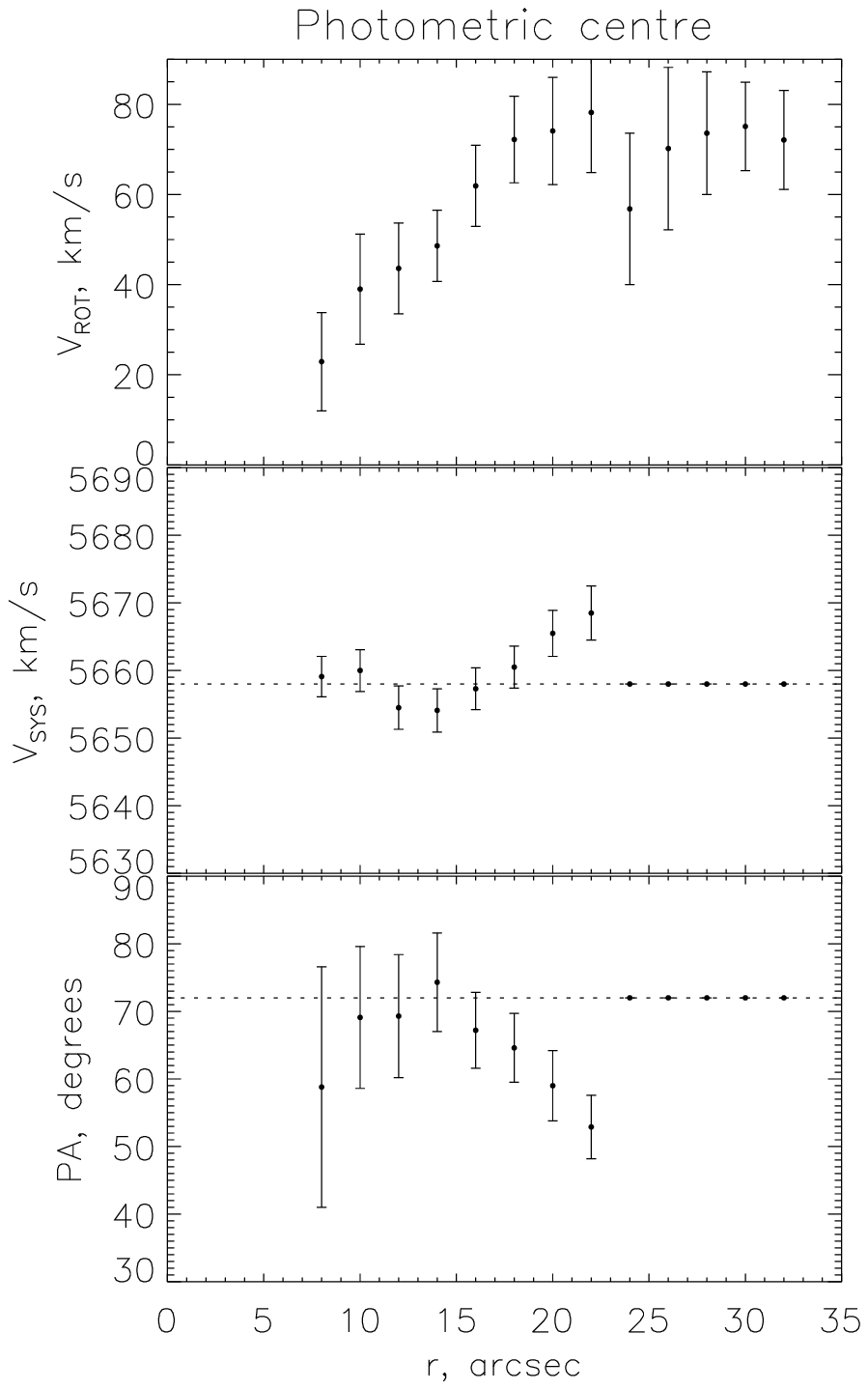}
\caption{The same as Fig.~\ref{fig_mod1}, but with the rotation centre coinciding with the photometric centre of the ring.}
\label{fig_mod2}
\end{figure}

Only points with velocity residuals  smaller than 20 km s$^{-1}$ were involved in the final fitting.  The parameters of circular rotation were calculated for each annular ring that was two arcsec wide. The radial dependencies of the rotation velocity (rotation curve), systemic velocity and $PA$ are presented on  Figs. \ref{fig_mod1} and \ref{fig_mod2} for the two above-mentioned models. The inclination $i$ was kept constant, corresponding to pure circular rotation models. We estimate  that the first model, using the kinematic centre, fits better since it has a slightly smaller amplitude of residual velocities (observed minus modeled) as well as smaller variations of $PA$ and $V_{sys}$ around their mean values. {\bf Among the attempts to model the kinematics of the object were also  fitting attempts including radial motions (expansion or contraction); from these we can put an upper limit to any expansion motion of 17 km s$^{-1}$.}

\subsection{Stellar populations}
\label{txt:model}

The determination of the stellar population compositions relies on the accurate modeling of the continua in the spectra. This is affected, in many cases, by  low S/N. Therefore, this modeling of individual knots is limited to the brighter knots and, even then, it may not be conclusive.


The stellar populations in each of the knots were derived by comparing the continuum spectral energy distribution with that predicted by models following Bruzual \& Charlot (2003; BC03) 
using the Starlight (SL) spectral synthesis code (Cid Fernandes et al. 2005).  SL finds a best-fitting overall spectral model as a superposition of the individual input ``base'' spectra and outputs the weights required from each individual model (separately for light and mass).
We found that the knots contain not only a very young stellar population, as could have been guessed from the strong emission lines, but also aged stellar populations.  The continuum fits described below were done using the BC03 models with the Padova 1993 tracks and a Salpeter IMF.  To assess uncertainties we used a Monte-Carlo method running each fitting hundreds of times while letting the observed spectrum change within its uncertainties.  Extinction was kept as a free parameter while the results below are from fits using a fixed Z=0.008 metallicity (0.4Z$_{\odot}$). However, we also ran the models where the metallicity was allowed to vary freely (in that case the best-fit metallicities become approximately 0.5Z$_{\odot}$), verifying that the results are within the uncertainties of about 0.15 dex in metallicity.

\begin{figure}
\centering{
  \includegraphics[width=16cm]{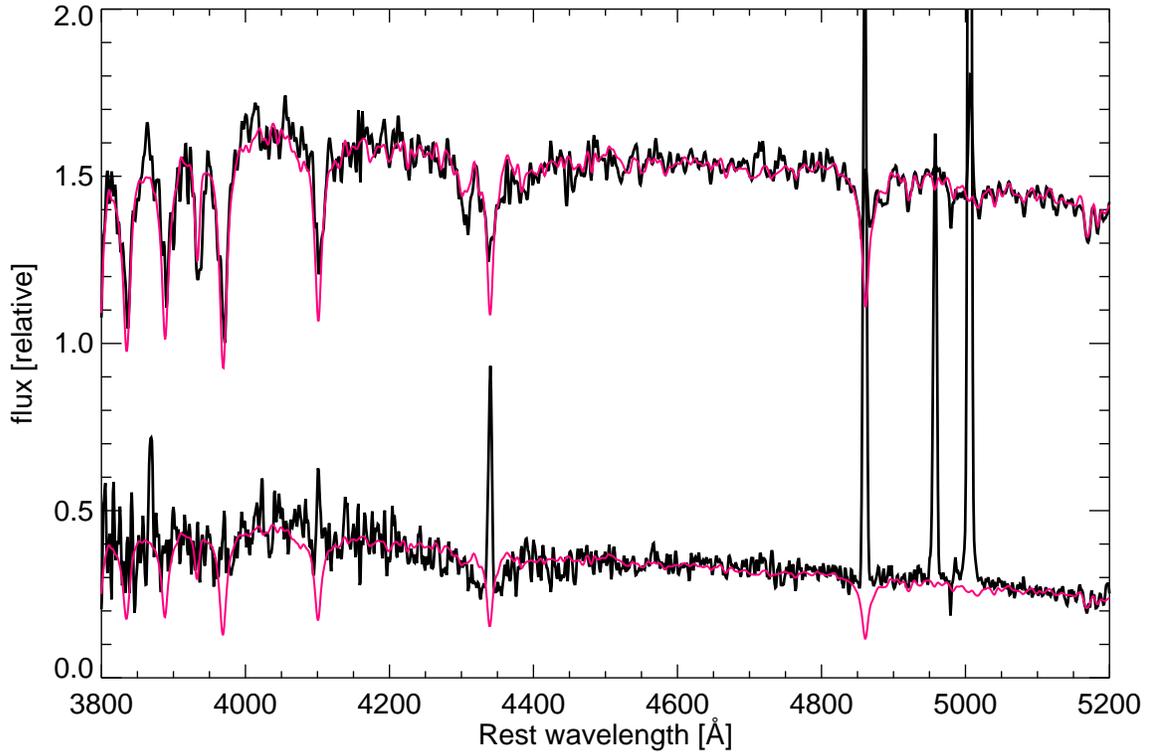}
  \caption{The observed extracted spectra of knots F (bottom) and I (top) are shown in black. The red curves overplot the best-fit BC03 models as found by Starlight (Table~3).    Note the much stronger Balmer and Calcium absorption features in I, while the former are essentially filled-in by emission lines in the F knot.  The sections with emission line contamination are ignored in the SL fitting.}
  \label{fig:slfits}}
\end{figure}

The specific results for individual knots originating from the BC03 models are listed in Table 3, assuming instantaneous delta-function star bursts, with an example of the fits in the most relevant spectral region for knots F and I being shown in Fig.~\ref{fig:slfits}. The various populations are numbered. It is clear that in addition to most knots showing strong current SF, all the knots contain various fractions of older stellar populations, especially an intermediate 150 Myr to 1 Gyr  population (while knot D fits result {\em only} in this range, the spectra do show emission lines there too indicating some current SF).   It is also clear that knot I is much older than the rest, with an underlying very old massive population not evident in others.  Though it appears the ring was not formed very recently, these SL fits do suggest a different history for it compared to the brightest (in continuum) knot I.

\begin{table}[th]
\label{t.models}
\begin{footnotesize}
\begin{center}
\vspace{0.5cm}
\begin{tabular}{|c|c|c|c|c|c|c|c|}
\hline
Knot &  age$_{light}$ (Myr) &  age$_{mass}$ (Gyr) & Pop 1 (\%) & Pop 2 (\%) & Pop 3 (\%) & Pop 4 (\%) & A$_V$ (mag) \\
\hline
A &  $130 \pm 30$   & $0.3 \pm 0.1$ & 37 &  3  &  55 &  7  &  0.8 \\
B &  $210 \pm 40$   & $0.4 \pm 0.1$ & 33 &  3  &  57 &  7  &  0.9 \\
D &  $670 \pm 80$   & $0.7 \pm 0.1$ & 0   &  0  & 100 & 0 &  0.0 \\
F &  $310 \pm 130$  & $1.6 \pm 1.6$ & 51 &  0  & 43 &  6  &  0.5 \\
I &   $790 \pm 110$  & $7.0 \pm 2.1$ & 0  &  22  & 68 &  9  &  0.6 \\
\hline
\end{tabular}
\caption{Stellar populations in the knots of ESO474 derived from the LS spectra with the BC03 models for delta-function star bursts.}
\end{center}
Notes to Table 3: Both {\em light} and {\em mass-averaged} ages are given in Columns 2 and 3,  the errors indicate the spread from running the SL fits hundreds of times while modifying the input spectrum randomly according to its uncertainty.  Columns 4 to 7 show percentages  of light coming from  stellar populations of ``current SF'', young, intermediate and old, defined as age ranges Pop 1: $<15$ Myr,  Pop 2: $15 < t < 150$ Myr,   Pop 3: $150 < t < 1000$ Myr  and Pop 4: $>1 Gyr$, respectively.  The uncertainty is a few percentage points in each case. The average derived extinction is listed in the last column with a typical error of $\pm 0.1$ mag.
\end{footnotesize}
\end{table}


We compared the model results for the youngest SF event against predictions derived from Starburst99 (SB99: Leitherer et al. 1999, 2014 and Vasquez et al. 2005) by measuring the equivalent widths of H$\alpha$.  SB99 predicts ages of 7 or 8 Myr for all the knots, from the best-fit BC03 models, except for knot I which shows a 16 Myr age. These are consistent with the results for the youngest component of the full SL fitting using BC03 models.

Since the spectra of knots I and F have higher S/N than the others, we attempted also extracting them in two sub-apertures each to see if any clear discontinuities, kinematic or stellar, would be evident.
The two parts of knot I show that the absorption lines fall pretty much on top of each other within the uncertainties of $\pm$30 km s$^{-1}$.  The two knot parts have the same stellar population and kinematics.
This is not entirely the case for knot F.  The SW and NE parts of knot F differ somewhat (the parts can be traced as visibly different knots e.g. in the the sum of FP scans of Fig.~\ref{fig:474b}).   The former part has significantly stronger emission lines, while the NW part towards knot I has signs of  somewhat older stellar populations.  However, we cannot disentangle the kinematics of the parts within our spectral resolution, i.e. we cannot see a difference in the absorption and emission wavelengths within the 30 km s$^{-1}$ uncertainty.


Using the WISE images, we measured knot F, the brightest one in line emission, with a 14" diameter aperture and found that its W1 and W2 magnitudes are 15.36$\pm$0.01 mag and 15.22$\pm$0.02 mag respectively. This knot produces a bit less than 20\% of the total emission from the galaxy in the two NIR bands.
The plots in Nikutta et al. (2014), in particular their Figure 4 where W1 is plotted against (W1--W2), show that knot F lies in the region of star-forming galaxies. This also confirms the indications from the LS spectra and from the models of the continuum in these spectra.

\subsection{Ionized gas}

Emission lines in the 1D spectra of ESO474 knots were measured using programs
described in Kniazev et al. (2004,2005). 
For each knot all intensities of emission lines relative to H$\beta$ [F($\lambda$)/F(H$\beta$)] and these ratios corrected for the extinction [I($\lambda$)/I(H$\beta$)] are shown in Table~\ref{t:Intens}.
The EW of the H$\beta$ emission is also listed along with the derived extinction coefficient $C$(H$\beta$) and $E(B-V)$ color excess.
The latter is a sum of the internal extinction in the target galaxy
and foreground extinction in the Milky Way.
The small total Milky Way extinction of A$_{\rm V}=$0.04~mag (Schelegel 1998)
suggests a non-negligible dust content of
A$_{\rm V} \sim 0.1-2.9$~mag extinction in the different knots of ESO474.
We note an interesting trend whereby the extinction is at its largest in the NE
part of the galaxy (knot I) and decreases towards the SW direction and is at its lowest at
knot D.

Our programs determine the
location of the continuum, perform a robust noise estimation, and fit
separate lines by a single Gaussian superimposed on the
continuum-subtracted spectrum. In this particular case,
since best-fit models were already created (see Section~\ref{txt:model} for details),
these models were used as the continuum and we did not use EW$_{abs}$($\lambda$)
for the hydrogen lines as a parameter in our calculations (e.g. Izotov et al. 1994).

The \ion{H}{ii} region spectrum was interpreted with the technique of
plasma diagnostics and iterative calculations
as described in detail in Kniazev et al. (2008). 
In our data the principal faint line [\ion{O}{iii}]$\lambda$4363 was
not detected thus we applied the so-called ``semi-empirical method'',
where the tight correlation between T(\ion{O}{iii}) and the total intensity of
the strong oxygen lines relative to I(H$\beta$) is used  
(Pagel et al. 1997, Shaver et al. 1983).
After T(\ion{O}{iii}) is estimated from the total intensity of
[\ion{O}{ii}]$\lambda$3727 and [\ion{O}{iii}]$\lambda\lambda$4959,5007 lines,
all other calculations were made as in the classic T$_{\rm e}$ method, using measured
line intensities.

Table~\ref{tab:Chem} presents the electron temperatures for
zones of emission of [\ion{O}{iii}] and  [\ion{O}{ii}],
ionic and total abundances of oxygen
and the electron number density $N_e$(\ion{S}{ii})
derived using the
[\ion{S}{ii}] $\lambda$6717/$\lambda$6731 lines ratio.
The minimum value of $N_e$(\ion{S}{ii}) was set to 10~cm$^{-3}$.
For the knot F both doublets [\ion{O}{ii}]$\lambda$3727, 3729 and
[\ion{O}{ii}]$\lambda$7320, 7330 were detected and we were able to
calculate $T_{\rm e}$(\ion{O}{ii})=8936$\pm$410\,K directly from these doublets ratios.
As seen from the above value and from Table~\ref{tab:Chem}, both $T_{\rm e}$(\ion{O}{ii}) for
knot F are equal within the formal errors and we may conclude that our T$_{\rm e}$
estimations are correct.

We found that the oxygen abundances in most of the knots (A--F)
did not differ within their rms uncertainties, with a weighted mean of
12+log(O/H)$\sim$8.52$\pm$0.07 dex.  This value is consistent, within uncertainties, with the stellar population metallicity of approximately half Solar, as derived from the full-spectrum fitting in the Section above.
Only knot I shows a strong difference in its oxygen abundance compared to the others, appearing to
be more metal-rich by approximately 0.5dex.

\begin{table*}[hbtp]
\centering{
\caption{Line intensities of the studied knots in ESO\,474$-$G040}
\label{t:Intens}
\begin{tabular}{lcccccc} \hline
\rule{0pt}{10pt}
& \MC{2}{c}{A knot} & \MC{2}{c}{B knot} & \MC{2}{c}{D knot} \\ \hline
\rule{0pt}{10pt}
$\lambda_{0}$(\AA) Ion &
F($\lambda$)/F(H$\beta$)&I($\lambda$)/I(H$\beta$) &
F($\lambda$)/F(H$\beta$)&I($\lambda$)/I(H$\beta$) &
F($\lambda$)/F(H$\beta$)&I($\lambda$)/I(H$\beta$) \\ \hline
3727\ [O\ {\sc ii}]\                           &  332.4$\pm$16.9 &  423.0$\pm$23.0 &  223.8$\pm$13.0 &  242.8$\pm$14.7 &  395.2$\pm$24.1 &  415.3$\pm$26.7    \\
4101\ H$\delta$\                               &   19.2$\pm$3.3  &   22.4$\pm$3.9  & ---             & ---             & ---             & ---                \\
4340\ H$\gamma$\                               &   28.8$\pm$3.5  &   31.9$\pm$3.9  &   23.4$\pm$2.5  &   24.3$\pm$2.6  &   12.2$\pm$1.5  &   12.5$\pm$1.5     \\
4861\ H$\beta$\                                &  100.0$\pm$6.5  &  100.0$\pm$6.5  &  100.0$\pm$5.6  &  100.0$\pm$5.6  &  100.0$\pm$6.7  &  100.0$\pm$6.7     \\
4959\ [O\ {\sc iii}]\                          &   76.7$\pm$5.4  &   75.3$\pm$5.3  &  103.3$\pm$5.1  &  102.7$\pm$5.0  &   56.5$\pm$4.9  &   56.3$\pm$4.9     \\
5007\ [O\ {\sc iii}]\                          &  243.4$\pm$12.0 &  237.0$\pm$11.8 &  299.3$\pm$12.4 &  296.7$\pm$12.3 &  151.7$\pm$8.6  &  150.9$\pm$8.6     \\
5876\ He\ {\sc i}\                             &    6.9$\pm$2.8  &    5.8$\pm$2.4  &    6.9$\pm$1.4  &    6.6$\pm$1.3  & ---             & ---                \\
6548\ [N\ {\sc ii}]\                           &   11.9$\pm$2.8  &    9.2$\pm$2.2  &   10.6$\pm$1.7  &    9.7$\pm$1.6  &   12.9$\pm$2.8  &   12.3$\pm$2.7     \\
6563\ H$\alpha$\                               &  377.2$\pm$17.6 &  290.8$\pm$14.8 &  318.2$\pm$13.0 &  291.4$\pm$12.9 &  307.5$\pm$15.2 &  291.5$\pm$15.7    \\
6584\ [N\ {\sc ii}]\                           &   37.9$\pm$3.2  &   29.1$\pm$2.6  &   33.3$\pm$2.4  &   30.5$\pm$2.2  &   40.8$\pm$3.3  &   38.7$\pm$3.3     \\
6717\ [S\ {\sc ii}]\                           &   69.0$\pm$4.8  &   52.2$\pm$3.8  &   34.3$\pm$2.4  &   31.2$\pm$2.3  &   60.5$\pm$4.5  &   57.2$\pm$4.4     \\
6731\ [S\ {\sc ii}]\                           &   52.5$\pm$4.4  &   39.7$\pm$3.4  &   27.4$\pm$2.3  &   24.9$\pm$2.1  &   41.5$\pm$3.9  &   39.2$\pm$3.8     \\
  & & & & & & \\
C(H$\beta$)\ dex          & \MC {2}{c}{0.34$\pm$0.06} & \MC {2}{c}{0.11$\pm$0.05} & \MC {2}{c}{0.07$\pm$0.06} \\
E(B-V)\ mag               & \MC {2}{c}{0.23$\pm$0.04} & \MC {2}{c}{0.07$\pm$0.03} & \MC {2}{c}{0.05$\pm$0.04} \\
EW(H$\beta$)\ \AA\        & \MC {2}{c}{  15$\pm$1}    & \MC {2}{c}{  21$\pm$1}    & \MC {2}{c}{  24$\pm$1}    \\
\hline
\rule{0pt}{10pt}
& \MC{2}{c}{F knot} & \MC{2}{c}{I knot} \\ \hline
\rule{0pt}{10pt}
3727\ [O\ {\sc ii}]\                           &  332.2$\pm$8.3 &  357.9$\pm$9.5  &  819.4$\pm$111.0 & 2226.0$\pm$331.6 \\
3868\ [Ne\ {\sc iii}]\                         &   16.4$\pm$2.7 &   17.5$\pm$2.9  & ---              & ---              \\
4101\ H$\delta$\                               &   12.3$\pm$0.8 &   12.9$\pm$0.9  & ---              & ---              \\
4340\ H$\gamma$\                               &   28.2$\pm$1.5 &   29.1$\pm$1.6  & ---              & ---              \\
4861\ H$\beta$\                                &  100.0$\pm$3.1 &  100.0$\pm$3.1  &  100.0$\pm$18.2  &  100.0$\pm$18.7  \\
4959\ [O\ {\sc iii}]\                          &   68.8$\pm$1.8 &   68.4$\pm$1.8  &  121.0$\pm$16.9  &  112.3$\pm$16.1  \\
5007\ [O\ {\sc iii}]\                          &  207.5$\pm$4.8 &  205.8$\pm$4.8  &  362.9$\pm$49.0  &  325.1$\pm$45.0  \\
5876\ He\ {\sc i}\                             &    9.4$\pm$0.5 &    9.0$\pm$0.5  & ---              & ---              \\
6300\ [O\ {\sc i}]\                            &   11.5$\pm$0.5 &   10.7$\pm$0.5  & ---              & ---              \\
6364\ [O\ {\sc i}]\                            &    3.5$\pm$0.4 &    3.2$\pm$0.3  & ---              & ---              \\
6548\ [N\ {\sc ii}]\                           &   12.8$\pm$0.4 &   11.8$\pm$0.4  &   72.6$\pm$13.5  &   24.9$\pm$4.9   \\
6563\ H$\alpha$\                               &  315.2$\pm$7.2 &  290.9$\pm$7.2  &  833.9$\pm$108.6 &  283.7$\pm$41.0  \\
6584\ [N\ {\sc ii}]\                           &   38.1$\pm$1.0 &   35.2$\pm$1.0  &  167.7$\pm$24.4  &   56.5$\pm$9.0   \\
6678\ He\ {\sc i}\                             &    3.0$\pm$0.4 &    2.7$\pm$0.4  & ---              & ---              \\
6717\ [S\ {\sc ii}]\                           &   57.9$\pm$1.5 &   53.2$\pm$1.5  &  333.9$\pm$55.8  &  105.1$\pm$19.1  \\
6731\ [S\ {\sc ii}]\                           &   40.5$\pm$1.2 &   37.2$\pm$1.2  &  256.4$\pm$49.7  &   80.2$\pm$16.6  \\
7320\ [O\ {\sc ii}]\                           &    3.2$\pm$0.4 &    2.9$\pm$0.3  & ---              & ---              \\
7330\ [O\ {\sc ii}]\                           &    2.0$\pm$0.3 &    1.8$\pm$0.3  & ---              & ---              \\
  & & & & \\
C(H$\beta$)\ dex          & \MC {2}{c}{0.10$\pm$0.03} & \MC {2}{c}{1.39$\pm$0.17} \\
E(B-V)\ mag               & \MC {2}{c}{0.07$\pm$0.03} & \MC {2}{c}{0.95$\pm$0.12} \\
EW(H$\beta$)\ \AA\        & \MC {2}{c}{  27$\pm$1}    & \MC {2}{c}{0.75$\pm$0.4}  \\
\hline
\end{tabular}
 }
\end{table*}

\begin{table*}[hbtp]
\centering{
\caption{Physical conditions and abundances in the studied knots of \ESO}
\label{tab:Chem}
\begin{tabular}{lccc} \hline
\rule{0pt}{10pt}
Value                                & A knot               & B knot               &  D knot               \\ \hline
$T_{\rm e}$(OIII)(K)\                &  9,077$\pm$1051~~    &  8,382$\pm$1043~~    &  8,549$\pm$1071~~     \\
$T_{\rm e}$(OII)(K)\                 &  9,420$\pm$678 ~~    &  9,323$\pm$1005~~    &  9,322$\pm$950 ~~     \\
$N_{\rm e}$(SII)(cm$^{-3}$)\         &  100$\pm$128~~       &  158$\pm$152~~       &  10$\pm$10 ~~         \\
O$^{+}$/H$^{+}$($\times$10$^5$)\     & 22.430$\pm$7.492~~   & 13.610$\pm$6.820~~   & 22.830$\pm$10.830~~   \\
O$^{++}$/H$^{+}$($\times$10$^5$)\    & 12.110$\pm$5.288~~   & 21.100$\pm$10.570~~  & 10.120$\pm$5.027~~    \\
12+log(O/H)\                         & ~8.54$\pm$0.12~~     & ~8.54$\pm$0.16~~     & ~8.52$\pm$0.16~~      \\
\hline
Value                                & F knot               & I knot               \\ \hline
$T_{\rm e}$(OIII)(K)\                &  8,533$\pm$1011~~    & 9,429$\pm$1329~~    \\
$T_{\rm e}$(OII)(K)\                 &  9,321$\pm$1357~~    & 9,970$\pm$296~~    \\
$N_{\rm e}$(SII)(cm$^{-3}$)\         &  10$\pm$10 ~~        & 104$\pm$345~~        \\
O$^{+}$/H$^{+}$($\times$10$^5$)\     & 17.920$\pm$5.238~~   & 89.750$\pm$18.170~~   \\
O$^{++}$/H$^{+}$($\times$10$^5$)\    & 13.490$\pm$6.327~~   & 11.990$\pm$5.672~~    \\
12+log(O/H)\                         & ~8.50$\pm$0.11~~     & ~9.01$\pm$0.08~~     \\
\hline
\end{tabular}
 }
\end{table*}

\section {Discussion}
\label{sec.discuss}

ESO474 has the appearance of an empty ring galaxy on sky survey images. We started this investigation in an attempt to understand its nature, origin and evolution. The results from the SALT observations show that the ring is not complete and uniform, but at least some of its elements (knots) contain considerable fractions, by mass, of old stellar populations while also forming young stars;  e.g., knot F is essentially very young. Each of the knots appears as a dwarf galaxy containing older stellar populations, while also undergoing present star  formation.


The deeper images show the presence of faint surface brightness features south of the brighter ring. The modeling of the stellar populations for those knots with reasonable S/N spectra indicates that all contain older stars, in most cases mixed together with very young stars. The entire ring complex fits a uniform circular rotation model, as shown by our FP and, to a lesser degree, by the LS data. 
 Considering the asymptotic velocity shown in the top-right plot of Fig.~\ref{fig_mod1} in the context of the Tully-Fisher relation for disks, the predicted absolute magnitude matches that calculated from the I-band photometry with the correct redshift. Despite its very peculiar shape, ESO474 follows well the TF relation.

After summarizing our findings, we now discuss various scenarios by which the object could acquire its present appearance. These were already mentioned in Section~\ref{sec:intro}: resonances in disks, galaxy collisions, and accretion of material from another galaxy or from the intergalactic space. None of these scenarios fits ESO474 perfectly.
A tidal interaction bringing in sufficient ISM to assure the star formation observed in ESO474 would have produced tidal tails and the donor galaxy would have been visible. A scenario based on a tidal interaction to form the apparent ring might have yielded an object similar to the Antennae (Arp 244) where the merger of two late-type galaxies with similar mass produced  long tails studded with star-forming clusters, but such long tails are not visible here. 

One possible analog, where objects are part of a ring-like structure in which star formation takes place, is the Leo Ring (Schneider et al. 1985, 1989). It is fair to note that the Leo Ring was discovered from a study of the extended HI distribution yet, as Thilker et al. (2009) have shown, massive star formation takes place in the ring in the form of (possibly) dwarf galaxies. Thilker et al. interpret the HI ring as primordial, and the dwarfs within the HI cloud as capable of survival by being relatively isolated from massive galaxies. Recently, Rosenberg et al. (2014) measured the metallitcity of the gas in the Leo ring via HST/COS spectra of QSOs in the background and found Z$\simeq$0.1Z$_{\odot}$, lower than expected from a tidal interaction between the large galaxies in the complex, but plausible if some mixing with low-Z, possibly primordial, material took place.

Another similar object is the galaxy (or group of galaxies) Klemola 25 (KL25: Danks \& Materne 1984). KL25 was ``speculated'' by Danks \& Materne to be the result of the breakout of a giant ring galaxy, possibly formed by a collision, via the bead-instability phenomenon. They also showed its similarity to a specific step in the Theys \& Spiegel (1977) simulation of a fragmenting ring. In contrast to ESO474, Klemola 25's four galaxies show essentially absorption-line spectra, as reported in Table 3 of Danks \& Materne.

An interpretation as a classical collisional RG would be convincing if a significant velocity difference would have been detected between the ring and either knot F or knot I, either of which could then be identified as the projectile. Our results for the line-of-sight velocity difference between the absorption and emission lines in and around knots I and F do not support this case, although a difference smaller than $\sim$20 km s$^{-1}$ would not have been detected. 
We note that knot F can be separated into two distinct clumps - though we see no kinematic differences within the uncertainties, the NE part towards knot I is more similar to the latter as far as its stellar population is concerned while the SW-part of knot F is the most strongly star forming of all the knots and has a spur attached to it faintly resembling a short tidal tail or a remnant of an arm. Hence, based on morphology alone, F-SW could feasibly be an intruder just incidentally superimposed on the ring currently.  However, this is not supported by any other evidence.  In contrast, knot I is much more clearly distinct from the other knots by its higher metallicity and older average stellar population.  However, since no kinematic difference was seen, more likely than it being an intruder would be this knot being a distinct morphological part of the original galaxy, such as an original nucleus.

This paper adds one more object to the set of early-type galaxies with outer rings of stars and gas. 
As the likeliest scenario, we propose that ESO474 is an old merger where the tails formed the ring as suggested by Moreno (2015) and Moreno et al. 2015, and that this ring broke up into the visible knots via the bead instability (Appleton \& Struck-Marcell 1996). Those beads, similar to dwarf galaxies, are now forming stars, possibly from material accreted from an extended disk that is detectable now as the ``fuzz'' south of the main ring. 
Given the metallicity difference and stellar population age difference of knot I compared to the rest of the ring, and the shift of the kinematic centre from the photometric one, it is possible that knot I could be a former galactic nucleus.

\section{Summary}
\label{sec.summary}
We presented new observations of the ``empty-ring'' galaxy ESO 474 G040 obtained at the Southern African Large Telescope using imaging with SALTICAM, and long-slit spectra and Fabry-P\'{e}rot interferometric imaging using the Robert Stobie Spectrograph. These observations show a ring in circular rotation around the kinematic centre. The knots in the ring are star-forming regions.  The lowest level of star-formation is seen in the brightest (in continuum light) knot I, which can be understand as one of the nuclei of two disk galaxies that merged sometime in the past. All the knots for which a population analysis could be performed show the presence of old stars in addition to young populations.  The knot suggested as an original nucleus has the oldest and most massive stellar population.  Based on recent hydro simulations, we understand the object as formed during a collision of two disk galaxies where the tidal tails gathered in the ring and are now forming stars.

\section{Acknowledgements}
The observations reported in this paper were obtained with the Southern African Large Telescope (SALT)
under programs 2012-2-RSA\_OTH-016, 2013-1-RSA\_OTH-020 and 2014-1-RSA\_OTH-010.   PV and AYK acknowledge support from the National Research Foundation of South Africa. The development of the FPI/SALT data analysis techniques was supported by the Russian Science Foundation (grant no. 14-22-00041). AM is also grateful for the financial support of the non-profit ``Dynasty'' Foundation. This research has made use of the NASA/IPAC Extragalactic Database (NED) which is operated by the Jet Propulsion Laboratory, California Institute of Technology, under contract with the National Aeronautics and Space Administration.




\end{document}